# A convolutional neural network for prestack fracture detection

*Zhenyu Yuan[*], China University of Petroleum - Beijing; Yuxin Jiang, PST Service Corporation; Jingjing Li, Beijing Concord O&M Wind Power Technology Co., Ltd.; Handong Huang, China University of Petroleum - Beijing.*

## Abstract

Fractures are widely developed in hydrocarbon reservoirs and constitute the accumulation spaces and transport channels of oil and gas. Fracture detection is a fundamental task for reservoir characterization. From prestack seismic gathers, anisotropic analysis and inversion were commonly applied to characterize the dominant orientations and relative intensities of fractures. However, the existing methods were mostly based on the vertical aligned facture hypothesis, it is impossible for them to recognize fracture dip. Furthermore, it is difficult or impractical for existing methods to attain the real fracture densities. Based on data-driven deep learning, this paper designed a convolutional neural network to perform prestack fracture detection. Capitalizing on the connections between seismic responses and fracture parameters, a suitable azimuth dataset was firstly generated through fracture effective medium modeling and anisotropic plane wave analyzing. Then a multi-input and multi-output convolutional neural network was constructed to simultaneously detect fracture density, dip and strike azimuth. The application on a practical survey validated the effectiveness of the proposed CNN model.

## Introduction

The recognition of fracture distributions is essential for characterizing reservoirs' storage capacity and production potential, and further guide well deployment and hydraulic fracturing.

Taking advantage of the anisotropic responses recorded by prestack seismic data, some methods were developed to recognize fractures. Through normal-moveout velocities calculating or shear-wave splitting analyzing, Tsvankin, (1997) and Li, (1999) presented some techniques (VVAz) to estimate fracture orientation and density. Based on azimuth anisotropy of seismic reflection amplitude, Gray and Head, (2000) performed amplitude variation with offset and azimuth (AVAz) analysis to determine fracture strike and density. The VVAz and AVAz techniques are either of low resolution or sensitive to noises. Rüger, (1998) derived P-wave reflectivity formula with offset and azimuth in HTI and orthorhombic media, and created the foundation for fracture inversion.. However, the inversion implementation required additional rock physics models to connect fracture parameters and elastic or anisotropic parameters. This process is heavily hypotheses based and impractical in realistic applications. Moreover, no matter deprivations for HTI or orthorhombic media, the fractures are assumed vertically aligned, it is impossible to detect real fracture dips.

Based on supervised deep learning, we proposed a convolutional neural network (CNN) to simultaneously detect fracture density, dip and orientation.

## Method

- Data Generation

The performance of data-driven deep learning heavily depends on the quality of prepared dataset. Azimuth seismic gathers provide a type of features for prestack fracture detection. Considering the seismic reflections are responses of strata interfaces, it is required to calculate some attributes to characterize the physical properties of certain reservoir formation. P-wave impedance is chosen as the azimuth feature here. However, the ground-truth labels including fracture density, dip or strike, are almost inaccessible to be directly observed. The formation microimager (FMI) image logging could private some interpretations of the fracture parameters, however, it is impractical to gather sufficient FMI logging images due to the high costs. Thus, we prepare the dataset through rock physics modelling.

The existence of fractures introduce anisotropy in rock medium. Linear slip model (Schoenberg, 1980) and Hudson crack model (Hudson, 1980) are adopted here to model the effective elastic properties of rocks containing vertically aligned fractures (HTI media). Considering the rotation of aligned fractures (TTI media), Bond transform is applied to perform coordinate transformation (Mavko et al., 2003).

$$\mathbf{C'} = \mathbf{MCM}^T, \qquad (1)$$

where $\mathbf{C}$ is stiffness matrix of HTI media, $\mathbf{C'}$ is stiffness matrix of rotated TTI media, $\mathbf{M}$ is the Bond transformation matrix, subscript $\mathbf{T}$ indicate matrix transposition.

Different from isotropic media, seismic waves propagate in anisotropic fractured media presents different velocities in different directions. Moreover, the shear wave split into two waves with different phase velocities and polarization directions, and coupled with the P wave. It is impossible to calculate the seismic wave velocities through analytical formulas. The Christoffel equation is appropriate for calculating velocities of the three types of waves propagating in various directions in anisotropic media (Dellinger, 1991). In accordance with the gathered prestack seismic gathers, only P wave is calculated and the propagation azimuths of seismic wave are set as 15°, 45°, 75°, 105°, 135° and 165°. Considering achievability and performance, the incidence of seismic wave is set as 40°. Multiplying P wave velocity and rock density, azimuth impedance gathers are generated.

# CNN for Prestack Fracture Detection

Table 1 Varying fracture parameters for dataset generation.

| Fracture Parameter Type | Variations |
|---|---|
| Density | 0.2, 0.1, 0.05, 0 |
| Dip (°) | [40,90], with step 10 |
| Strike Azimuth (°) | [0,180), with step 10 |

Taking fracture density, dip and strike azimuth as 3 target labels, whose ground-truth values are varied in ranges showed in Table 1. To consider the spreading extend of fractures and enhance the data stability, the azimuth impedance feature is prepared as an 8*6 array, some samples are shown in Figure 1.

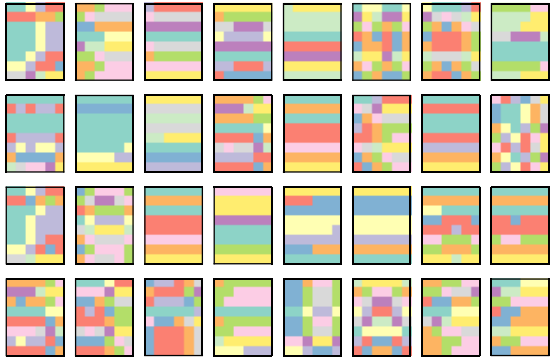

Figure 1 Some samples of prepared azimuth P-wave impedance.

The azimuth values of wave propagation direction are associated with the azimuth P-wave impedances, providing another type of feature, expressed as a 1*6 vector.

- Network Architecture

Taking 8*6 array impedances and 1*6 vector azimuths as features and 3 fracture parameters as labels, prestack fracture detection deep learning is a task with multi inputs and multi outputs. Based on the general hybrid deep neural network architecture (Yuan et al., 2020), a multi-input multi-output convolutional neural network is designed, shown as Figure 2.

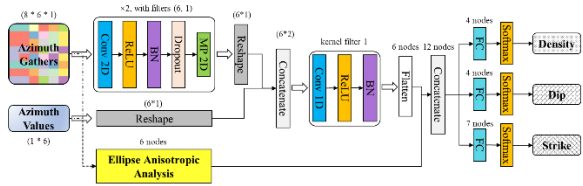

Figure 2 Architecture of proposed prestack fracture detection CNN model.

More than complete data-driven learning, the proposed architecture takes a physical process into consideration. The azimuth impedances of anisotropic fractured media vary with wave azimuth as an ellipse, and the ellipse parameters like center location, lengths of the semi-major axis and semi-minor axis, rotation angle, are closely relevant to fracture density, dip and strike. For feature representation, ellipse anisotropic analysis is performed to extract 6 geometric features, namely center location $x_0$ and $y_0$, lengths of semi-major axis and semi-minor axis, relative ratio of semi-major and semi-minor axis, rotation angle.

For 8*6 azimuth impedance feature, two coterminous 2D convolution units are designed to perform feature learning. Each convolution unit contains a 2D convolution filter, a ReLU activation function, a batch normalization layer, a dropout layer and a 2D max pooling layer. Then the learned features are concatenated with the input azimuth vector to construct a 1D sequence with 2 channels. The new 1D sequence are then processed by a 1D convolution unit to extract high-hierarchy features. At last, the features extracted from structural 2D and 1D inputs are concatenated with ellipse geometric features to perform final target learning. For three fracture labels, three fully-connected branch networks with softmax activation function are used to perform separate multi-class classification.

- Model Training

The proposed CNN model establishes the mapping relation between 2 inputs and 3 outputs, expressed as

$$(Y^1, Y^2, Y^3) = F(\mathbf{X}^1, \mathbf{X}^2), \qquad (2)$$

where $\mathbf{X}^1$ and $\mathbf{X}^2$ indicates input azimuth value vector and impedance array, $Y^1$, $Y^2$ and $Y^3$ indicates three fracture labels. $F$ is an abstract function of the CNN model, representing the combination of different network units and the process of ellipse anisotropic analysis.

For separate multi-class classification, the cross-entropy measure is taken as loss function,

$$CE^i = -\sum_{j=1}^{M}\sum_{k=1}^{C^i} Y_{jk}^i \log(p_{jk}^i) \qquad (3)$$

where $i$ indicates index of classification task, varying from 1 to 3, $k$ indicates class index, varying from 1 to $C^i$ for each classification task. $Y_{jk}$ indicates binary indicator of class $k$ for instance $j$, $p_{jk}$ indicates predicted probability of class $k$ for instance $j$. The calculation of probability is subject to the selection of $F$.

The total loss of prestack seismic detection model is expressed as the sum of three separate classification branch,

$$CE = \sum_{i=1}^{3} CE^i \qquad (4)$$

# CNN for Prestack Fracture Detection

To ensure accuracy and generalization of the CNN model, fracture dips from 40° to 90° are binned into three categories as 45°, 65° and 85°, azimuths are binned into six categories as 10°, 40°, 70°, 100°, 130°, and 160°. In addition, when the fracture density is 0, the corresponding fracture dip and azimuth are invalid, set as None.

The prepared dataset is randomly split as training, validation and test data by ratios of 65%, 15% and 20%. During model training, both training and validation losses and accuracies are calculated to demonstrate the training performance. Figure 3 shows the total loss of the CNN model, where blue and orange lines show training and validation performances separately. Both the training and validation loss curves monotonically decreases with iterations, demonstrating that the model training achieve a pleasant convergence.

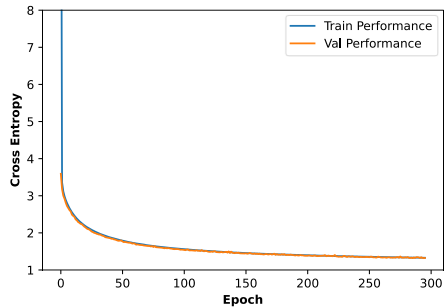

Figure 3 Training loss and validation loss varying with iterations.

Figure 4 to Figure 6 demonstrate training and validation accuracy for fracture density, dip and azimuth respectively. On the whole, with the increasing of iterations, the training and validation accuracy increase rapidly in the early stage and then stretch stably. The validation accuracy curve fluctuate around the training accuracy curve. In general, the CNN model achieves an acceptable accuracy for three fracture parameters. Detection of fracture density shows good performance with accuracy above 90%, the accuracies of fracture dip and azimuth detection are about 80%.

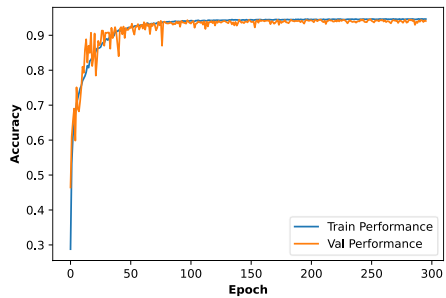

Figure 4 Training and validation accuracy for fracture density.

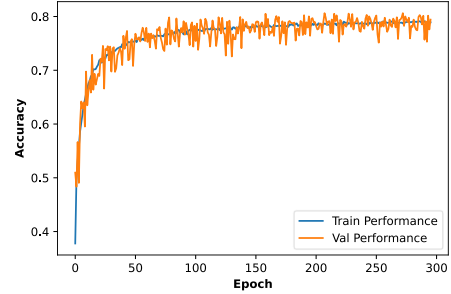

Figure 5 Training and validation accuracy for fracture dip.

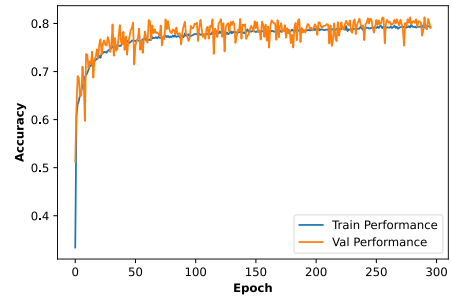

Figure 6 Training and validation accuracy for fracture strike azimuth.

- Model Evaluation

Confusion matrix is a useful metric to evaluate performance of multi-class classification, providing straightforward visualization of instance numbers of correct and mistaken classes. Applying the well-trained CNN model to prepared test data, Figure 7 to Figure 9 demonstrate confusion matrixes for fracture density, dip and strike azimuth respectively. Consistent with the accuracy evaluation, the performance of fracture density detection is the best, accurately recognize the density range. For fracture dip and azimuth, the main diagonal of confusion matrix accounts for the majority of the predictions, representing that the fracture attitudes could be properly detected with a high probability.

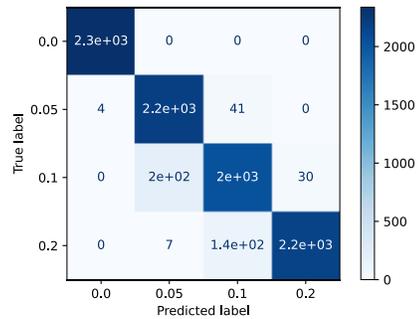

Figure 7 Confusion matrix of test data for fracture density.

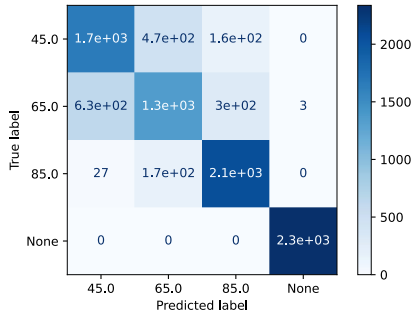

Figure 8 Confusion matrix of test data for fracture dip.

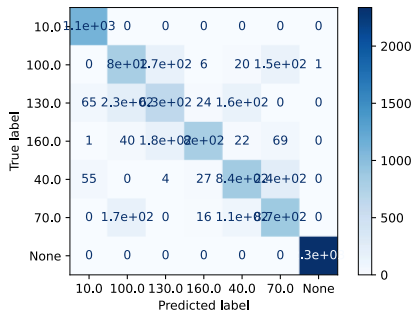

Figure 9 Confusion matrix of test data for fracture strike azimuth.

## Examples

Taking a coalbed methane reservoir as example, the well-trained CNN model is adopted to the prestack azimuth gathers to detect fracture parameters. The target reservoir formation develops north-east and north-north-east folds and faults, as well as extensive high-steep fractures.

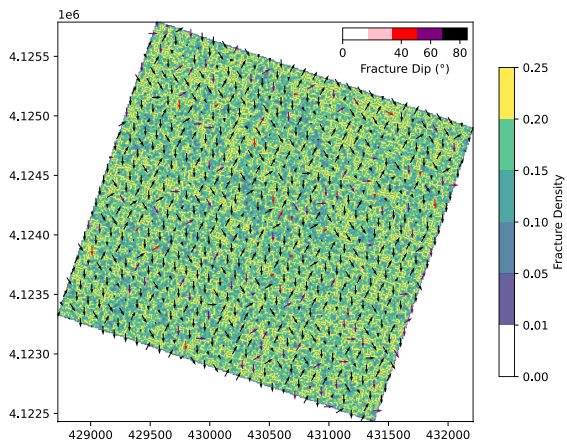

Figure 10 Prestack fracture detection by proposed CNN model.

Figure 10 shows the fracture detection result by the proposed CNN model. The background image indicates fracture density, the direction, length and color of arrows separately indicate strike orientation, density and dip angle of fractures. It can be concluded that high-steep fractures are widely developed in the whole survey, exhibiting a dominant orientation in the north-north-east direction. Without image-logging wells, a discontinuity attribute (Yuan et al., 2019) is calculated as a relevant reference, shown as Figure 11. It shows that the fracture density and orientation are consistent with the distribution of faults.

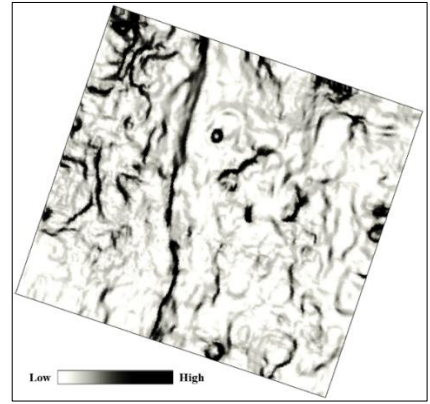

Figure 11 Fault detection by seismic discontinuity attribute.

## Conclusions

Driven by generated azimuth dataset, a convolutional neural network was constructed to perform prestack fracture detection. Two different types of data, namely 2D matrix-type azimuth gathers and 1D vector-type azimuth values, were taken as input features. In addition, three fracture parameters including density, dip angle and strike azimuth were taken as target labels. Thus, the proposed CNN is a hybrid network with multi inputs and multi outputs. Furthermore, the proposed network takes full advantage of the azimuth dataset's features of spatial variation and geometric shape. More than pure data-driven procedure, some geometric features contained in azimuth variation were extracted through ellipse anisotropic analyzing. The model training demonstrates that the CNN model is of pleasant accuracy and generalization. Applied to a practical survey, the well-trained neural network model achieved reasonable inference consistent with geologic conclusions. Comparing with commonly used fracture detection methods, the proposed network show advantages of simultaneously and quantitatively detecting fracture density, dip and azimuth. To further improve the performance, the proposed CNN model can be generalized by bringing in some fault attributes as additional features to introduce more efficient constraints.

# CNN for Prestack Fracture Detection

**References**

Dellinger, J. A., 1991, Anisotropic seismic wave propagation, Stanford University.

Gray, D., and K. Head, 2000, Fracture detection in Manderson Field: A 3-D AVAZ case history: The Leading Edge, **19**, no. 11, 1214-1221.

Hudson, J. 1980, Overall properties of a cracked solid. Paper read at Mathematical Proceedings of the Cambridge Philosophical Society

Li, X.-Y., 1999, Fracture detection using azimuthal variation of P-wave moveout from orthogonal seismic survey lines: Geophysics, **64**, no. 4, 1193-1201.

Mavko, G., T. Mukerji, and J. Dvorkin, 2003, The rock physics handbook: Tools for seismic analysis in porous media: Cambridge University Press.

Rüger, A., 1998, Variation of P-wave reflectivity with offset and azimuth in anisotropic media: Geophysics, **63**, no. 3, 935-947.

Schoenberg, M., 1980, Elastic wave behavior across linear slip interfaces: The Journal of the Acoustical Society of America, **68**, no. 5, 1516-1521.

Tsvankin, I., 1997, Reflection moveout and parameter estimation for horizontal transverse isotropy: Geophysics, **62**, no. 2, 614-629.

Yuan, Z., H. Huang, Y. Jiang, and J. Li, 2020, Hybrid Deep Neural Networks for Reservoir Production Prediction: Journal of Petroleum Science and Engineering, 108111. http://dx.doi.org/https://doi.org/10.1016/j.petrol.2020.108111.

Yuan, Z., H. Huang, Y. Jiang, J. Tang, and J. Li, 2019, An Enhanced Fault Detection Method based on Adaptive Spectral Decomposition and Super-Resolution Deep Learning: Interpretation, **7**, no. 3, T713–T725.